\documentclass[11pt]{article}
\usepackage[utf8]{inputenc}      
\usepackage[T1]{fontenc}         
\usepackage{amsmath, amssymb}    
\usepackage{amsfonts, mathrsfs}  
\usepackage{amsthm}              
\usepackage{bm}                  
\usepackage{graphicx}            
\usepackage{color}               
\usepackage{cite}                
\usepackage{enumerate}           
\usepackage{geometry}            
\usepackage{float}               
\usepackage{ifthen}              
\usepackage[caption=false]{subfig} 
\usepackage{epstopdf}            
\usepackage{braket}              
\usepackage{booktabs}
\usepackage[breaklinks=true]{hyperref}            

\begin{document}

\title{\bf Average metric adjusted skew information of coherence under conical 2-designs generalized equiangular measurements}

\vskip0.1in
\author{\small Baolong Cheng$^1$, Linlin Ye$^1$, Zhaoqi Wu$^1$\thanks{Corresponding author. E-mail: wuzhaoqi\_conquer@163.com}\\
{\small\it  1. Department of Mathematics, Nanchang University,
Nanchang 330031, P R China}}

\date{}
\maketitle

\noindent {\bf Abstract} {\small }\\
Quantum coherence is an important quantum resource which plays a
pivotal role in the field of quantum information. Based on metric
adjusted skew information, we define a measure of quantum
uncertainty to study average coherence under conical 2-designs
generalized equiangular measurements, and prove the equivalence of
this measure to the scaled average coherence based on metric
adjusted skew information under a set of unitary groups, operator
orthonormal bases, and mutually unbiased bases. We also derive two
trade-off relations by this measure and solve a conjecture.
Furthermore, we give two entanglement criteria by this measure and
conical 2-designs generalized equiangular measurement, respectively,
and illustrate the effectiveness of them by explicit examples.


\vskip0.2in

\noindent {\bf 1. Introduction}\par Quantum coherence, as a
fundamental and prominent feature in quantum mechanics, serves as a
critical resource in quantum information science
\cite{DickeRH}. It is not only a
description of quantum wave such as entanglement, interference, and
diffraction, but also the cause of quantum interference, quantum
nonlocality, and quantum entanglement \cite{FanY2025}. With the
development of quantum information science, the quantification of
coherence has emerged as a central problem. In a landmark work,
Baumgratz et al.\cite{Baumgratz} introduced a quantification scheme
for coherence resource, while Yu et al. \cite{YuX} proposed an
equivalent framework, which may be easier to verify in certain
situations.

The Wigner-Yanase (WY) skew information \cite{WignerEP} and the
Wigner-Yanase-Dyson (WYD) skew information \cite{LiebEH} have been
introduced respectively. After that, the WYD skew information was
further extended to the generalized Wigner-Yanase-Dyson (GWYD) skew
information \cite{ChenP}. Finally, these concepts were unified
within metric adjusted skew information \cite{HansenF}, and its
properties were explored intensively \cite{CaiL}. On the other hand,
the non-Hermitian version of WY skew information and WYD skew
information (modified skew information) was proposed in
\cite{DouY2013JMP} and \cite{DouY2014IJTP}, respectively. The
uncertainty relations for various kinds of generalized skew
information
\cite{DouY2013JMP,DouY2014IJTP,WuZQ2020IJTP,HuangHJ2020EPL,XuC2024CTP}
and metric adjusted skew information
\cite{XuC2024QIP,TangL2026AQT}
have been investigated in recent years.

Coherence measure based on skew information has been proposed in
\cite{GirolamiD2014PRL,YuCS2017PRA}, providing a fundamental insight
into the relationship between quantum states and observables, and
playing a key role in understanding uncertainty within quantum
systems \cite{LuoS2017PRA,SunY2021PRA}. Based on skew information,
average coherence \cite{LuoS2019PLA,WuZQ2021JPA,FanY2023PRA},
coherence generating power \cite{WuZQ2022QIP}, geometric features
\cite{WuZQ2020CTP} and uncertainty relations \cite{ShengYH2021QIP}
have been discussed. Moreover, skew information of coherence with
respect to channels
\cite{LuoS2018PRA,WuZQ2020QIP}, and
uncertainty of channels via skew information
\cite{XuC2022QIP,XuC2024PS} have been explored. Notably,
metric adjusted skew information has also been used to quantify
quantum correlations \cite{RenRN2024QIP,FanY2024MPLA} and
nonclassicality \cite{TangL2025PRA}.

Quantum measurement directly influences how the coherence in quantum
states are manifested \cite{GriffithsB}. As a generalized quantum
measurement method, positive operator-valued measures (POVMs)
overcomes the restrictions of traditional projective measurements,
offering more flexible measurement schemes \cite{BischofF2019}. In
the framework of POVMs, mutually unbiased bases (MUBs)
\cite{Schwinger} and symmetric informationally complete measurements
(SIC-POVMs) \cite{Renes} are key concepts which are closely related
\cite{Renes,Bengtsson}. To overcome the limitations imposed
by the rank-one elements of SIC-POVMs and MUBs, Kalev
\cite{KalevA2014} introduced mutually unbiased measurements (MUMs)
and Appleby \cite{ApplebyDM2007} proposed general symmetric
informationally complete measurements (GSIC-POVMs), enabling
effective quantum measurements across a broader range of dimensions.
In order to establish a unified understanding between MUMs and
GSIC-POVMs, the concept of $(N,M)$-POVMs \cite{SiudzinskaK} was
presented, which was later generalized to the generalized
equiangular measurements (GEAMs) \cite{SiudzinskaK2024}. Recent
studies have investigated the average metric adjusted skew
information of coherence over unitary groups\cite{FanY2023PRA},
orthonormal operator bases\cite{FanY2023PRA}, MUBs
\cite{FanY2023PRA}, MUMs\cite{ChenB} and GSIC-POVMs \cite{ChenB}
were studied, while the average GWYD skew information of coherence
over $(N,M)$-POVMs have been analyzed \cite{ChenRX}. Based on
quantum measurements, several entanglement criteria were proposed
\cite{TangL2025PRA,ChenB,TangL2025}.

The remainder of this paper is structured as follows. In Section 2,
we recall some preliminary concepts. The main results and some
corollaries are presented in Section 3. We give two
criteria of entanglement detection and some corollaries in Section
4. Some concluding remarks are given in Section 5.

\vskip0.1in

\noindent {\bf 2. Preliminaries}\par

Throughout this paper, denote by $\mathcal{H}$ the $d$-dimensional
Hilbert space, $\mathcal{D}(\mathcal{H})$ the set of density
matrices (quantum states) on $\mathcal{H}$,
$\mathcal{A}=\{\ket{i}\}_{i=1}^{d}$ the reference basis of
$\mathcal{H}$, $\mathcal{P}=\mathrm{tr}\rho^2$ the purity of quantum
state $\rho$, $\mathbb{B}_\rho$ the basis composed by the
orthonormal eigenvectors of $\rho$, $\mathbb{I}_d$ the $d\times d$
identity matrix, and $\mathbb{R}$ the set of real numbers.

\noindent{\bf 2.1 Metric adjusted skew information}\par

We begin with the concept of metric adjusted skew information
\cite{HansenF}, which is a general family of skew information
defined by
\begin{align}
\label{eq1}
I_f(\rho, H) = \frac{f(0)}{2} \mathrm{tr}\left\{ \mathrm{i}[\rho, H] \, c_f(L_\rho, R_\rho) \, \mathrm{i}[\rho, H] \right\},
\end{align}
where $c_f(x,y)=[yf(xy^{-1})]^{-1}$ is called the Morozova-Chentsov
function, $f$ is a positive operator monotone function satisfying
$f(x)=xf(x^{-1})$, $f(1)=1$ and $f(0)=\lim\limits_{x\to0}f(x)>0$,
and the superoperators $L_\rho$ and $R_\rho$ are defined by
$L_\rho(H)=\rho H$ and $R_\rho(H)=H \rho$ respectively. The metric
adjusted skew information has many useful properties
\cite{HansenF,CaiL,TakagiR}:

(1) $I_f(\rho, H) = 0$, if $[\rho, H] = 0$;

(2) $I_f\left(\sum\limits_k p_k \rho_k, H\right) \leq \sum\limits_k p_k I_f(\rho_k, H)$, where $p_k \geq 0$ and $\sum\limits_k p_k = 1$;

(3) $I_f(\rho_A \otimes \rho_B, H_{AB}) = I_f(\rho_A, H_A) + I_f(\rho_B, H_B)$, where $H_{AB} = H_A \otimes I_B + I_A \otimes H_B$.

If one considers
\begin{align}
\notag
f_{\alpha,\beta}(x)=\frac{2\alpha\beta(x-1)^2}{(x^\alpha-1)(x^\beta-1)(x^{1-\alpha-\beta}+1)}, x>0,
\end{align}
where $\alpha, \beta\ge0$ and $\alpha+\beta\le1$, the corresponding
Morozova-Chentsov function is
\begin{align}
\label{eq2} c_{\alpha,\beta}(x,y) = \frac{(x^\alpha -
y^\alpha)(x^\beta - y^\beta)\big( x^{1-\alpha-\beta} +
y^{1-\alpha-\beta} \big)}{2\alpha\beta\, (x-y)^2},
\end{align}
and $I_f(\rho, H)$ in Eq.\eqref{eq1} becomes the GWYD skew
information $I_{\alpha,\beta}(\rho, H)$. Specifically, the WYD skew information corresponds to $f_{\alpha,1-\alpha}(x)=\frac{\alpha(1-\alpha)(x-1)^2}{(x^\alpha-1)(x^{1-\alpha}-1)}$, and the WY skew information is recovered by taking $\alpha=\beta=\frac{1}{2}$, for which $f_{\frac{1}{2},\frac{1}{2}}(x)=\left(\frac{1+\sqrt{x}}{2}\right)^2$.


Gibilisco formulated a general representation of quantum uncertainty
based on metric adjusted skew information \cite{GibiliscoP},
\begin{align}
\notag
Q_{f}(\rho) = \sum_{i=1}^{d^{2}} I_{f}(\rho, H_{i}),
\end{align}
which is basis-independent \cite{CaiL2018}. Moreover, let
$\rho=\sum\limits_{j=1}^{d}\lambda_j\ket{\phi_j}\bra{\phi_j}$ be the
spectral decomposition of a quantum state $\rho$, then we have
\begin{align}
\label{eq3}
Q_{f}(\rho) = \frac{f(0)}{2}\sum_{m,n = 1}^{d}(\lambda_m-\lambda_n)^2c_{f}(\lambda_m,\lambda_n).
\end{align}

If one considers the GWYD skew information, the quantity in Eq.\eqref{eq3} becomes $Q_{\alpha,\beta}(\rho)$. Furthermore, if one considers the WYD skew information, $Q_{\alpha,\beta}(\rho)$ becomes $Q_\alpha(\rho)$.

\noindent{\bf 2.2 GEAMs}\par

To make the notation clear from the outset, we summarize in Table 1 the main parameters used in the construction of GEAMs.

\begin{table}[H]
\centering
\caption{Summary of the main parameters of GEAMs.}
\begin{tabular}{c p{0.9\linewidth}}
\toprule
Symbol & Description \\
\midrule
$\mathcal{P}$ & Generalized equiangular measurement (GEAM) \\
$\mathcal{P}_k$ & $k$-th frame of $\mathcal{P}$ \\
$P_{k,l}$ & $l$-th measurement operator in the $k$-th frame \\
$M_k$ & Number of operators in the $k$-th frame \\
$\gamma_k$ & Normalization factor \\
$a_k$ & Trace of $P_{k,l}$ \\
$b_k$ & Purity of the normalized $P_{k,l}$ \\
$c_k$ & Overlap between different measurement operators within the same frame \\
$\tau_k$ & Real parameter controlling the deviation of $P_{k,l}$ from $\frac{a_k}{d}\mathbb{I}_d$ \\
$S_k$ & Half of the squared Frobenius distance between different measurement operators within the same frame, characterizing the separation between measurement operators in the frame \\
$S$ & Same separation between measurement operators within each frame for conical 2-designs GEAMs \\
\bottomrule
\end{tabular}
\end{table}

A GEAM $\mathcal{P}=\bigcup\limits_{k=1}^{N}\mathcal{P}_k$ is a
collection of $N$ generalized equiangular tight frames
$\mathcal{P}_k=\{ P_{k,l}: l=1,\cdots,M_k \}$ such that
\cite{SiudzinskaK2024}

(1) the elements of a single frame sum up to $\sum\limits_{l=1}^{M_k}P_{k,l} = \gamma_k \mathbb{I}_d$ with a probability distribution $\{ \gamma_k \}$;

(2) the total number of operators $\mathcal{P}_{k,l}$ is bounded by $\sum\limits_{k=1}^{N}M_k=d^2+N-1$;

(3) they satisfy the following trace conditions
\begin{align}
&\mathrm{tr}(P_{k,l}) = a_k, \notag\\
&\mathrm{tr}(P^2_{k,l}) = b_k \mathrm{Tr}(P_{k,l})^2, \notag\\
&\mathrm{tr}(P_{k,l}P_{k,l^\prime}) = c_k \mathrm{Tr}(P_{k,l})\mathrm{Tr}(P_{k,l^\prime}), \quad l\neq l^\prime, \notag\\
&\mathrm{tr}(P_{k,l}P_{k^\prime,l^\prime}) =
r\mathrm{Tr}(P_{k,l})\mathrm{tr}(P_{k^\prime,l^\prime}), \quad k\neq
k^\prime,\notag
\end{align}
where the symmetry parameters
\begin{align}\notag
a_{k} = \frac{d\gamma_{k}}{M_{k}}, \quad
c_{k} = \frac{M_{k} - db_{k}}{d(M_{k} - 1)}, \quad
r = \frac{1}{d}, \quad
\frac{1}{d} < b_{k} \leq \frac{1}{d} \min\{d, M_{k}\}.
\end{align}
Now let us consider the construction of GEAMs \cite{SiudzinskaK2025}. In this quantum measurement framework, the systematic construction employs an orthonormal Hermitian basis such that
\begin{align}
\notag
\left\{G_0=\frac{\mathbb{I}_d}{\sqrt{d}},G_{k,l}:k=1,\cdots,N;l=1,\cdots,M_k-1;\mathrm{tr}G_{k,l}=0\right\}.
\end{align}
The measurement operators are defined as\cite{SiudzinskaK2024}
\begin{align}
\notag
P_{k,l}=\frac{a_k}{d}\mathbb{I}_d+\tau_k H_{k,l}
\end{align}
with a real parameter
\begin{align}
\notag \tau_{k} = \pm \sqrt{\frac{S_{k}}{M_{k}(\sqrt{M_{k}} +
1)^2}},\,\, S_{k} = a_{k}^2 (b_{k} - c_{k}),
\end{align}
where $H_{k,l}$ is a family of traceless operators satisfying
\begin{align}
\notag
H_{k,l}=\begin{cases}
  G_k-\sqrt{M_k}(\sqrt{M_k}+1)G_{k,l}, & l=1,\cdots,M_k-1, \\
  (\sqrt{M_k}+1)G_k, & l=M_k,
\end{cases}
\end{align}
with $G_k=\sum\limits_{l=1}^{M_k-1}G_{k,l}$. Chosen in such a way,
it is claimed that $P_{k,l}$ are positive operators.

In Table 2, we restate the popular examples of the generalized
equiangular measurements together with the explicit values of their
defining parameters given in \cite{SiudzinskaK2025}. Note that the
constants corresponding to the families of quantum measurements were
rescaled because they had to sum up to the identity operator.
\begin{table}[H]
\centering
\begin{tabular}{c c c c c c}
\toprule
 & MUBs & MUMs & SIC POVM & GSIC POVM & $(N,M)$-POVM \\
\midrule
$N$ & $d+1$ & $d+1$ & $1$ & $1$ & $N$ \\

$M_k$ & $d$ & $d$ & $d^{2}$ & $d^{2}$ & $M$ \\

$\gamma_k$ & $\frac{1}{d+1}$ & $\frac{1}{d+1}$ & $1$ & $1$ & $\frac{1}{N}$ \\

$a_k$ & $\frac{1}{d+1}$ & $\frac{1}{d+1}$ & $\frac{1}{d}$ & $\frac{1}{d}$ & $\frac{d}{NM}$ \\

$b_k$ & $1$ & $\frac{1}{d}<b\leq 1$ & $\frac{1}{d}$ & $\frac{1}{d}<b\leq 1$ & $\frac{1}{d}<b\leq\frac{1}{d}\min\{d,M\}$ \\

$c_k$ & $0$ & $\frac{1-b}{d-1}$ & $\frac{1}{d+1}$ & $\frac{d-b}{d^{2}-1}$ & $\frac{M-db}{d(M-1)}$ \\

$S$ & $\frac{1}{(d+1)^{2}}$ & $\frac{db-1}{(d+1)(d^{2}-1)}$ & $\frac{1}{d(d+1)}$ & $\frac{db-1}{d(d^{2}-1)}$ & $\frac{d(db-1)}{NM(d^{2}-1)}$ \\
\bottomrule
\end{tabular}
\caption{ \cite{SiudzinskaK2025} Parameters characterizing GEAMs for popular classes of quantum measurements.}
\end{table}

Specifically, letting $S_k=S$ for all $k=1,\cdots,N$ with $0\le S
\le\min\left\{\frac{d\gamma_{k}^2}{M_{k}},\frac{d-1}{M_{k}-1}\frac{d\gamma_{k}^2}{M_{k}}\right\}$,
we obtain a subclass of GEAMs which is called conical 2-designs
GEAMs \cite{SiudzinskaK2025}. In what follows, our results will
focus on this special kind of GEAMs.

Conical 2-designs GEAMs provide a unified framework encompassing MUBs, MUMs, SIC-POVMs, GSIC-POVMs, and ($N$,$M$)-POVMs, making the results obtained under this framework broadly applicable. Moreover, arbitrary POVMs, including GEAMs, are experimentally realizable via linear-optical schemes \cite{MedendorpZED} or Naimark-dilation techniques on superconducting platforms \cite{IvashkovP}. Thus the study of conical 2-designs GEAMs is both theoretically significant and experimentally relevant.

\vskip0.1in

\noindent {\bf 3. The uncertainty relations of metric adjusted skew information}

In this section, we discuss the uncertainty relations of metric adjusted skew information of coherence.

We first define a measure of quantum uncertainty with respect to conical 2-designs GEAMs based on
the metric adjusted skew information as
\begin{align}
\label{eq4}
C_{f}^{\mathcal{P}}(\rho) = \frac{1}{N} \sum_{k = 1}^{N} \sum_{l = 1}^{M_k} I_{f}(\rho, P_{k, l}),
\end{align}
where $\mathcal{P}=\{ P_{k,l}: l=1,\cdots,M_k, k=1,\cdots,N \}$ is a
GEAM. Then we have the following result.

{\bf Theorem 1} The average coherence of a state $\rho$ under conical 2-designs GEAM in $\mathcal{H}$ can be expressed as
\begin{align}
\label{eq5}
C_{f}^{\mathcal{P}}(\rho)=\frac{S}{N}Q_{f}(\rho),
\end{align}
where $0\le S \le \min\left\{\frac{d\gamma_{k}^2}{M_{k}},\frac{d-1}{M_{k}-1}\frac{d\gamma_{k}^2}{M_{k}}\right\}$.

The detailed proof of Theorem 1 is given in Appendix A.

By fixing different chosen parameters and Morozova-Chentsov
functions, many consequent results can be derived from Theorem 1.
For $\frac{1}{d}<b\le1$, if $N=d+1$ and
$S=\frac{db-1}{(d+1)(d^2-1)}$, a GEAM reduce to a MUM, the average
coherence in Eq.\eqref{eq4} becomes $C_{f}^{\mathrm{MUM}}(\rho)$,
and if $N=1$, $S=\frac{db-1}{d(d^2-1)}$, a GEAM reduce to a
GSIC-POVM, the average coherence in Eq.\eqref{eq4} becomes
$C_{f}^{\mathrm{GSIC-POVM}}(\rho)$. Then from Theorem 1, we have the
following corollary, in which (1) and (2) are equivalent to Theorem
1 and Theorem 2 in \cite{ChenB}, respectively.

{\bf Corollary 1} (1) The average coherence of a state $\rho$ under MUMs in $\mathcal{H}$ can be expressed as
\begin{align}
\notag
C_{f}^{\mathrm{MUM}}(\rho)=\frac{db-1}{(d+1)(d^2-1)}Q_{f}(\rho);
\end{align}

(2) The average coherence of a state $\rho$ under GSIC-POVMs in $\mathcal{H}$ can be expressed as
\begin{align}
\notag
C_{f}^{\mathrm{GSIC-POVM}}(\rho)=\frac{db-1}{d(d^2-1)}Q_{f}(\rho),
\end{align}
where $0\le S \le \min\left\{\frac{d\gamma_{k}^2}{M_{k}},\frac{d-1}{M_{k}-1}\frac{d\gamma_{k}^2}{M_{k}}\right\}$.

For the Morozova-Chentsov function $c_{\alpha,\beta}(x,y)$ in
Eq.\eqref{eq2}, the average coherence in Eq.\eqref{eq4} becomes
$C_{\alpha,\beta}^{\mathcal{P}}(\rho)$, then we have the following
corollary.

{\bf Corollary 2} The average coherence of a state $\rho$ under conical 2-designs GEAMs in $\mathcal{H}$ can be expressed as
\begin{align}
\notag
C_{\alpha,\beta}^{\mathcal{P}}(\rho)=\frac{S}{N}Q_{\alpha,\beta}(\rho),
\end{align}
where $0\le S \le \min\left\{\frac{d\gamma_{k}^2}{M_{k}},\frac{d-1}{M_{k}-1}\frac{d\gamma_{k}^2}{M_{k}}\right\}$.

\textbf{Remark 1} For $b_k=b$, where $\frac{1}{d}<b\le1$, when
$S=\frac{d(db-1)}{NM(d^2-1)}$, a conical 2-designs GEAM reduces to a
$(N,M)$-POVM, then Corollary 2 reduces to Theorem 1 in
\cite{ChenRX}. Moreover, if $\alpha+\beta=1$, for
$\frac{1}{d}<b\le1$, $S=\frac{d(db-1)}{NM(d^2-1)}$, Corollary 2 is
equivalent to Theorem 4 in \cite{SiudzinskaK2025}.

In fact, we can construct a generalized symmetric measurement
$\widetilde{\mathcal{P}}=\bigcup\limits_{k=1}^{N}\frac{\mathcal{P}_k}{\gamma_k}$
via a GEAM \cite{SiudzinskaK2024}. Thus we can define a measure of
quantum uncertainty with respect to a generalized symmetric
measurement corresponding to a conical 2-designs GEAM based on the
metric adjusted skew information as
\begin{align}
\notag
C_{f}^{\widetilde{\mathcal{P}}}(\rho) = \frac{1}{N} \sum_{k = 1}^{N} \sum_{l = 1}^{M_k} I_{f}\left(\rho, \frac{P_{k, l}}{\gamma_k}\right),
\end{align}
where $\gamma_k=\frac{1}{N}$ for all $k=1,\cdots,N$. Then we have
the following corollary.

{\bf Corollary 3} The average coherence of a state $\rho$ under a generalized symmetric measurement corresponding to conical 2-designs GEAMs in $\mathcal{H}$ can be expressed as
\begin{align}
\label{eq6}
C_{f}^{\widetilde{\mathcal{P}}}(\rho)=NSQ_{f}(\rho),
\end{align}
where $0\le S \le \min\left\{\frac{d\gamma_{k}^2}{M_{k}},\frac{d-1}{M_{k}-1}\frac{d\gamma_{k}^2}{M_{k}}\right\}$.

\textit{Proof} Let $\gamma_k=\frac{1}{N}$ for all $k=1,\cdots,N$.
Then it holds that
\begin{align}
\label{eq7} C_{f}^{\widetilde{\mathcal{P}}}(\rho) = \frac{1}{N}
\sum_{k = 1}^{N} \sum_{l = 1}^{M_k} I_{f}(\rho, \frac{P_{k,
l}}{\gamma_k}) = N^2 C_{f}^{\mathcal{P}}(\rho).
\end{align}
Combining Eq.\eqref{eq5} with Eq.\eqref{eq7}, we have
Eq.\eqref{eq6}.\qed

This corollary means that metric adjusted skew information has a
greater average coherence on the generalized symmetric measurements
than on the generalized equiangular measurements. Although the
latter has greater parametric freedom \cite{SiudzinskaK2024}, it
does not result in greater average coherence.

Now let us consider the average coherence over all orthonormal bases
defined as \cite{FanY2023PRA}
\begin{align}
\notag
C_f^{\mathcal{U}}(\rho)=\int_{\mathcal{U}}C_{f}^{U\Pi U^\dagger}(\rho)\mathrm{d}U,
\end{align}
where $\mathrm{d}U$ denotes the normalized Haar measure on the full
unitary group $\mathcal{U}$ of the system Hilbert space, and $U\Pi U^\dagger=\left\{ U \ket{i}\bra{i} U^\dagger: i=1,\cdots,d \right\}$. Then we have the following corollary.

{\bf Corollary 4} For any $\rho\in \mathcal{D}(\mathcal{H})$ and any Morozova-Chentsov function $f$, we have
\begin{align}
\label{eq8}
C_{f}^{\mathcal{P}}(\rho) = \frac{S(d+1)}{N}C_f^{\mathcal{U}}(\rho),
\end{align}
where $0\le S \le \min\left\{\frac{d\gamma_{k}^2}{M_{k}},\frac{d-1}{M_{k}-1}\frac{d\gamma_{k}^2}{M_{k}}\right\}$.

\textit{Proof} By Proposition 1 in \cite{FanY2023PRA}, it holds that
\begin{align}
\label{eq11} C_{f}^{\mathcal{\mathcal{U}}}(\rho) = \frac{1}{d+1}Q_{f}(\rho).
\end{align}
Combining Eq.\eqref{eq5} with Eq.\eqref{eq9}, we obtain
Eq.\eqref{eq8}. \qed

\textbf{Remark 2} By Proposition 2 in \cite{FanY2023PRA}, for any
$\rho\in \mathcal{D}(\mathcal{H})$ and any Morozova-Chentsov
function $f$, we have
\begin{align}
\notag C_f^{\mathcal{U}}(\rho) = C_f^{\mathrm{OB}}(\rho),
\end{align}
where $C_f^{\mathrm{OB}}(\rho)$ denote the average coherences over
elements of any operator orthonormal basis. Then we have the
following equivalence (up to a constant factor),
\begin{align}
\notag C_{f}^{\mathcal{P}}(\rho) =
\frac{S(d+1)}{N}C_f^{\mathcal{U}}(\rho) =
\frac{S(d+1)}{N}C_f^{\mathrm{OB}}(\rho).
\end{align}
Furthermore, if one considers quantum state of any prime power
dimensional system, by Proposition 4 in \cite{FanY2023PRA}, it holds
that
\begin{align}
\notag C_f^{\mathcal{U}}(\rho) = C_f^{\mathrm{MUB}}(\rho),
\end{align}
where $C_f^{\mathrm{MUB}}(\rho)$ denote the average coherence over
any complete family of MUBs. Thus we have the following equivalence
(up to a constant factor) in prime power dimensional systems,
\begin{align}
\notag C_{f}^{\mathcal{P}}(\rho) =
\frac{S(d+1)}{N}C_f^{\mathcal{U}}(\rho) =
\frac{S(d+1)}{N}C_f^{\mathrm{OB}}(\rho) =
\frac{S(d+1)}{N}C_f^{\mathrm{MUB}}(\rho).
\end{align}

Let $C_{f}(\rho) = C_f^{\mathcal{U}}(\rho) = C_f^{\mathrm{OB}}(\rho)
= C_f^{\mathrm{MUB}}(\rho)$. Then the corresponding quantum $f$
entropy $S_{f}(\rho)$ is defined by \cite{FanY2023PRA}
\begin{align}
\label{eq10}
S_{f}(\rho) = d-1 - (d+1)C_{f}(\rho),
\end{align}
and the quasientropy is defined by \cite{PetzD1985}
\begin{align}
\notag S_{f}^{X_k}(\rho_1|\rho_2)=\mathrm{tr}(X_k^\dagger
c_{f}^{-1}(L_{\rho_1}, R_{\rho_2}) X_k),
\end{align}
where $\{X_k\}_{k=1}^{d^2}$ is the set of operator orthonormal
basis. Letting $\tilde{f}(x) = \frac{1}{2} \left[ (x + 1) - (x -
1)^2 \frac{f(0)}{f(x)} \right]$, the connections between quantum $f$
entropy and quasientropy is\cite{FanY2023PRA}
\begin{align}
\label{eq11} S_{f}(\rho)=
\sum\limits_{k=1}^{d^2}S_{\tilde{f}}^{X_k}(\rho|\rho) - 1.
\end{align}
Combining Corollary 4 and \eqref{eq10}, we can easily obtain the
following theorem.

{\bf Theorem 2} The trade-off relation between quantum $f$ entropy
and average coherence is
\begin{align}
\label{eq12}
S_{f}(\rho) + \frac{N}{S}C_{f}^{\mathcal{P}}(\rho) = d-1.
\end{align}

Furthermore, combining \eqref{eq11} and \eqref{eq12}, we have the
following corollary.

{\bf Corollary 5} The trade-off relation between quasientropy and
average coherence is
\begin{align}
\label{eq13}
\sum\limits_{k=1}^{d^2}S_{\tilde{f}}^{X_k}(\rho|\rho) + \frac{N}{S}C_{f}^{\mathcal{P}}(\rho) = d.
\end{align}

These two trade-off relations indicate that the entropy of a state cannot be arbitrarily small when the average coherence is very small.

Letting $N=d+1$, $S=\frac{1}{(d+1)^2}$ and $N=1$,
$S=\frac{1}{d(d+1)}$ in Eq.\eqref{eq4} respectively, the average
coherence in Eq.\eqref{eq4} becomes $C_{f}^{\mathrm{MUB}}(\rho)$ and
$C_{f}^{\mathrm{SIC-POVM}}(\rho)$ respectively. For the
Morozova-Chentsov function $c_{\alpha,\beta}(x,y)$ in
Eq.\eqref{eq2}, $C_{f}^{\mathrm{MUB}}(\rho)$ and
$C_{f}^{\mathrm{SIC-POVM}}(\rho)$ become
$C_{\alpha,\beta}^{\mathrm{MUB}}(\rho)$ and
$C_{\alpha,\beta}^{\mathrm{SIC-POVM}}(\rho)$ respectively. In
\cite{ChenRX}, Chen et al. proved the following result
\begin{align}
\notag
C_{\alpha,\beta}^{\mathrm{MUB}}(\rho)+C_{\alpha,\beta}^{\mathrm{SIC-POVM}}(\rho)=I^{\max}_{\alpha,\beta}(\rho),
\end{align}
where $\alpha,\beta\ge0$, $\alpha+\beta\le1$ and
$I^{\max}_{\alpha,\beta}(\rho)$ is the maximal coherence of a state
based on $c_{\alpha,\beta}(x,y)$, and they conjectured that \cite{ChenRX} this result retains its validity within the broader framework of quantum coherence.

Let $I^{\max}_f(\rho):=\max\limits_{\Pi}
I_f(\rho,\Pi)$ be the maximal coherence of a state based on the
metric adjusted skew information, where the maximum is taken over
all von Neumann measurements $\Pi$\cite{FanY2025}. We resolve the
conjecture by giving the following result.

{\bf Theorem 3} For any $\rho\in \mathcal{D(H)}$ and any
Morozova-Chentsov function $f$, we have
\begin{align}
\label{eq14}
C_{f}^{\mathrm{MUB}}(\rho)+C_{f}^{\mathrm{SIC-POVM}}(\rho)=I^{\max}_f(\rho).
\end{align}

The detailed proof of Theorem 3 is given in Appendix B.

Note that the above three quantities are all defined
in terms of the metric adjusted skew information, which includes the
GWYD skew information as a special case. Thus Theorem 3 presents a
result in a broader framework, thereby solving the conjecture in
\cite{ChenRX}.

\noindent {\bf 4. Entanglement detection via
$C_{f}^{\mathcal{P}}(\rho)$ and conical 2-designs GEAMs} \par
Consider bipartite states
$\rho_{AB}\in\mathcal{D}(\mathcal{H_A}\otimes\mathcal{H_B})$,
$\mathcal{P}_{A}=\{P_{k,l}^A\}$ and $\mathcal{P}_{B}=\{P_{k,l}^B\}$
be any two sets of conical 2-designs GEAMs with the same parameters
on $\mathcal{H}_A$ and $\mathcal{H}_B$, respectively. Define
\begin{align}
\label{eq15}
F_{f}^{\mathcal{P}_A,\mathcal{P}_B}(\rho_{AB})=\frac{1}{N} \sum_{k = 1}^{N} \sum_{l = 1}^{M_k} I_{f}(\rho_{AB}, P_{k, l}^A\otimes \mathbb{I}_B + \mathbb{I}_A\otimes P_{k, l}^B).
\end{align}
Now we establish a criterion of entanglement detection as follows.

{\bf Theorem 4} The state $\rho_{AB}$ is entangled if
\begin{align}
\label{eq16}
F_{f}^{\mathcal{P}_A,\mathcal{P}_B}(\rho_{AB})>\frac{2S(d-1)}{N},
\end{align}
where $0\le S \le \min\left\{\frac{d\gamma_{k}^2}{M_{k}},\frac{d-1}{M_{k}-1}\frac{d\gamma_{k}^2}{M_{k}}\right\}$.

The detailed proof of Theorem 4 is given in Appendix C.

\textbf{Remark 3} For a generalized symmetric measurement
corresponding to conical 2-designs GEAMs
$\widetilde{\mathcal{P}}^A=\{\frac{P_{k,l}^A}{\gamma_{k}}\}$ and
$\widetilde{\mathcal{P}}^B=\{\frac{P_{k,l}^B}{\gamma_{k}}\}$, with
$\gamma_k=\frac{1}{N}$ for all $k=1,\cdots,N$, it holds that
\begin{align}
\notag
\widetilde{F}_{f}^{\widetilde{\mathcal{P}}^A,\widetilde{\mathcal{P}}^B}(\rho_{AB})=\frac{1}{N} \sum_{k = 1}^{N} \sum_{l = 1}^{M_k} I_{f}\left(\rho_{AB}, \frac{P_{k, l}^A\otimes \mathbb{I}_B + \mathbb{I}_A\otimes P_{k, l}^B}{\gamma_k}\right)=N^2 F_{f}^{\mathcal{P}^A,\mathcal{P}^B}(\rho_{AB}),
\end{align}
then the state $\rho_{AB}$ is entangled if
\begin{align}
\notag
\widetilde{F}_{f}^{\widetilde{\mathcal{P}}^A,\widetilde{\mathcal{P}}^B}(\rho_{AB})>2NS(d-1).
\end{align}

According to Theorem 4, for $b_k=b$, where $k=1,\cdots,N$, $\frac{1}{d}<b\le1$, letting $N=d+1$, $S=\frac{db-1}{(d+1)(d^2-1)}$ and $N=1$, $S=\frac{db-1}{d(d^2-1)}$, conical-2 design GEAM $\mathcal{P}^A$, $\mathcal{P}^B$ reduce to MUM $\mathcal{P}_{\mathrm{MUM}}^A$, $\mathcal{P}_{\mathrm{MUM}}^B$ and GSIC-POVM $\mathcal{P}_{\mathrm{GSIC-POVM}}^A$, $\mathcal{P}_{\mathrm{GSIC-POVM}}^B$, then we have the following corollary in which (1) and (2) are equivalent to Theorem 3 and Theorem 4 in \cite{ChenB}, respectively.

{\bf Corollary 6} Let $b_k=b$, $k=1,\cdots,N$, $\frac{1}{d}<b\le1$.
Then we have

(1) The state $\rho_{AB}$ is entangled if
\begin{align}
\notag
F_{f}^{\mathcal{P}_{\mathrm{MUM}}^A,\mathcal{P}_{\mathrm{MUM}}^B}(\rho_{AB})>\frac{2(db-1)}{(d+1)^3};
\end{align}

(2) The state $\rho_{AB}$ is entangled if
\begin{align}
\notag
F_{f}^{\mathcal{P}_{\mathrm{GSIC-POVM}}^A,\mathcal{P}_{\mathrm{GSIC-POVM}}^B}(\rho_{AB})>\frac{2(db-1)}{d(d+1)^2}.
\end{align}

Inspired by \cite{TangL2025}, consider another quantity to establish
a criterion of entanglement detection. Consider the bipartite state
$\rho_{AB}\in\mathcal{D}(\mathcal{H_A}\otimes\mathcal{H_B})$,
$\rho^A(\rho^B)$ stands for the reduced density matrix of the
subsystem $A(B)$, $\mathcal{P}^A=\{P_{k,l}^A\}$ and
$\mathcal{P}^B=\{P_{k,l}^B\}$ are any two sets of conical 2-designs
GEAM with the same parameters on $\mathcal{H}_A$ and
$\mathcal{H}_B$, respectively. Define
\begin{align}
\label{eq17} G^{\mathcal{P}^A,\mathcal{P}^B}(\rho_{AB})=\sum_{k =
1}^{N}\sum_{l = 1}^{M_k} \left| \mathrm{tr}((P_{k, l}^A \otimes
P_{k, l}^B)(\rho_{AB}-\rho_A\otimes\rho_B)) \right|.
\end{align}
Then we have the following theorem.

{\bf Theorem 5} The state $\rho^{AB}$ is entangled if
\begin{align}
\label{eq18} G^{\mathcal{P}^A,\mathcal{P}^B}(\rho_{AB})>
S\sqrt{(1-\mathrm{tr}\rho_A^2)(1-\mathrm{tr}\rho_B^2)},
\end{align}
where $0\le S \le \min\left\{\frac{d\gamma_{k}^2}{M_{k}},\frac{d-1}{M_{k}-1}\frac{d\gamma_{k}^2}{M_{k}}\right\}$.

The detailed proof of Theorem 5 is given in Appendix D.

\textbf{Remark 4} For a generalized symmetric measurement corresponding to conical 2-designs GEAMs $\widetilde{\mathcal{P}}^A=\{\frac{P_{k,l}^A}{\gamma_{k}}\}$ and $\widetilde{\mathcal{P}}^B=\{\frac{P_{k,l}^B}{\gamma_{k}}\}$, where $\gamma_k=\frac{1}{N}$ for all $k=1,\cdots,N$, it holds that
\begin{align}
\notag
\widetilde{G}^{\widetilde{\mathcal{P}}^A,\widetilde{\mathcal{P}}^B}(\rho_{AB})=\sum_{k
= 1}^{N}\sum_{l = 1}^{M_k} \left| \mathrm{tr}\left(\left(\frac{P_{k,
l}^A}{\gamma_k} \otimes \frac{P_{k,
l}^B}{\gamma_k}\right)(\rho_{AB}-\rho_A\otimes\rho_B)\right)
\right|=N^2 G^{\mathcal{P}^A,\mathcal{P}^B}(\rho_{AB}),
\end{align}
then the state $\rho_{AB}$ is entangled if
\begin{align}
\notag
\widetilde{G}^{\widetilde{\mathcal{P}}^A,\widetilde{\mathcal{P}}^B}(\rho_{AB})> N^2S\sqrt{(1-\mathrm{tr}\rho_A^2)(1-\mathrm{tr}\rho_B^2)}.
\end{align}

Now we give two examples to show the validity and efficiency of the
criteria in the above two theorems.

{\bf Example 1} Consider the $d$-dimensional
isotropic states \cite{LiN}
\begin{align}
\notag \rho^{\rm iso}_{AB} = \frac{1-q}{d^2} \mathbb{I}_{AB} +
q\ket{\phi^+}\bra{\phi^+}, \quad 0 \le q \le 1,
\end{align}
where $\ket{\phi^+} = \frac{1}{\sqrt{d}} \sum\limits_{i=1}^d \ket{i}
\otimes \ket{i}$. Take a set of conical 2-designs GEAMs
$\mathcal{P}^A=\{P_{k,l}^A\}$, and let $\mathcal{P}^B=\{P_{k,l}^B\}$
denote the conjugation of $\mathcal{P}^A$. It is obvious that they
have the same index sets. For the isotropic states $\rho^{\rm
iso}_{AB}$, a direct calculation using the spectral decomposition
$\rho^{\rm
iso}_{AB}=\lambda_1|\phi^+\rangle\langle\phi^+|+\lambda_2(\mathbb{I}_{AB}-|\phi^+\rangle\langle\phi^+|)$
with $\lambda_1=\frac{1-q+qd^2}{d^2}$ and
$\lambda_2=\frac{1-q}{d^2}$ yields
\[
I_{f}(\rho^{\rm iso}_{AB}, P_{k, l}^A\otimes \mathbb{I}_B +
\mathbb{I}_A\otimes P_{k, l}^B) =
f(0)q^2c_f\!\left(\lambda_1,\lambda_2\right)\cdot\frac{4S(M_k-1)}{dM_k}.
\]
If $f(t)=\frac{t+1}{2}$, then $f(0)=\frac12$ and
$c_f(x,y)=\frac{2}{x+y}$. Substituting these into the above
expression and summing over $k,l$, we have
\[
F_{f}^{\mathcal{P}^A,\mathcal{P}^B}(\rho^{\rm iso}_{AB}) =
\frac{4q^2S(d^2-1)d}{N[2(1-q)+qd^2]}.
\]
Utilizing Theorem 4, by simple calculations, we
obtain
\begin{align}
\notag
2d(d+1)q^2 - (d^2 - 2)q - 2 > 0.
\end{align}
Therefore, if $q>\frac{d^2-2+\sqrt{(d^2-2)^2+16d(d+1)}}{4d(d+1)}$,
$\rho_{\mathrm{iso}}$ must be entangled and conforms to previous
results in \cite{ChenB}. It is easy to verify that
$\frac{d^2-2+\sqrt{(d^2-2)^2+16d(d+1)}}{4(d+1)}>\frac{d^2-2+\sqrt{(d^2-2)^2+16d(d+1)}}{4d(d+1)}>\frac{1}{d+1}$,
which means that we have a better criterion than the one in
\cite{CaiL2018} but a weaker criterion than the one in
\cite{HuangF,HorodeckiM}.

{\bf Example 2} Consider the $d$-dimensional Werner
states \cite{LiN}
\begin{align}
\notag \rho^{\rm w}_{AB} = \frac{d-x}{d^3-d} \mathbb{I}_{AB} +
\frac{dx-1}{d^3-d} F, \quad -1 \le x \le 1,
\end{align}
where $F$ is the swap operator defined as
$F=\sum\limits_{i,j=0}^{d-1}\ket{ij}\bra{ij}$. Take two sets of
conical 2-designs GEAMs $\mathcal{P}^A=\{P_{k,l}^A\}$ and
$\mathcal{P}^B=\{P_{k,l}^B\}$ that are the same. For the Werner
states $\rho^{\rm w}_{AB}$, it holds that
$\rho_A=\rho_B=\frac{\mathbb{I}_d}{d}$. Since
$\operatorname{tr}((P_{k,l}^A\otimes
P_{k,l}^B)F)=\operatorname{tr}(P_{k,l}^A P_{k,l}^B)=b_k a_k^2$, we
have
\[
\operatorname{tr}((P_{k,l}^A\otimes P_{k,l}^B)\rho^{\rm w}_{AB}) =
\frac{d-x}{d^3-d}a_k^2 + \frac{dx-1}{d^3-d}b_k a_k^2,
\]
and
\[
\operatorname{tr}((P_{k,l}^A\otimes P_{k,l}^B)(\rho_A\otimes\rho_B))
= \frac{a_k^2}{d^2}.
\]
Thus
\[
\operatorname{tr}((P_{k,l}^A\otimes P_{k,l}^B)(\rho^{\rm
w}_{AB}-\rho_A\otimes\rho_B)) = a_k^2\frac{(b_k
d-1)(dx-1)}{d^2(d^2-1)}.
\]
Noting that $\sum\limits_{k,l}a_k^2(b_k d-1)=Sd(d^2-1)$ holds for
conical 2-designs GEAMs, we obtain
\[
G^{\mathcal{P},\mathcal{Q}}(\rho^{\rm w}_{AB}) = \frac{S|1-dx|}{d}.
\]
Therefore, if $-1\le x <\frac{2}{d}-1$, $\rho^{\rm
w}_{AB}$ must be entangled by Theorem 5 and this result is
consistent with \cite{TangL2025, ShenSQ,ZhangCJ}. Compared with the
PPT criterion \cite{PeresA,HorodeckiM1996}, which detects the
entanglement of Werner states for $x<0$ in all dimensions, our
criterion is weaker. The reduction criterion \cite{HorodeckiM}
detects entanglement for $x<-\frac{1}{d}$, which is weaker than
ours, as it fails to detect entanglement in Werner states for
$d\ge3$.
\begin{figure}[H]
\centering
\includegraphics[width=0.65\textwidth]{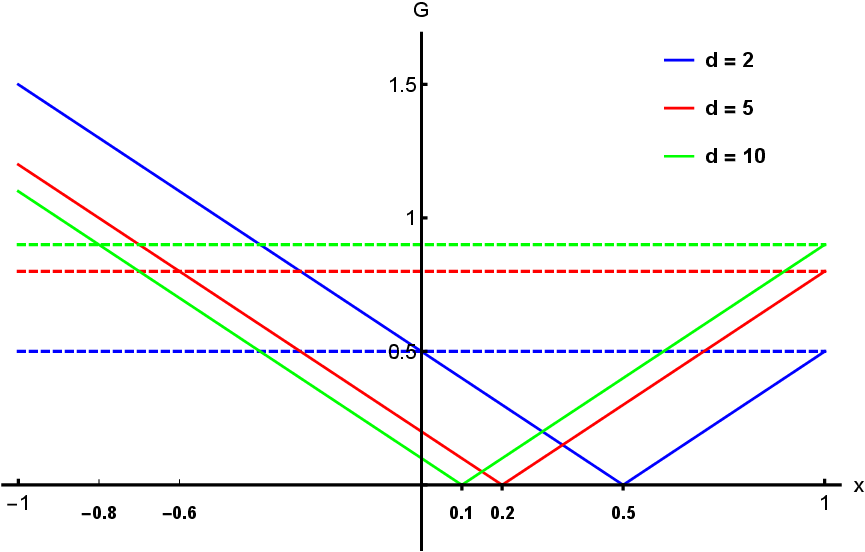}
\caption{Plot of $G^{\mathcal{P},\mathcal{Q}}(\rho_{AB})$ (solid lines) and its threshold $\frac{S(d-1)}{d}$ (dashed lines) as functions of $x$ for Werner states with $d=2,5,10$ (blue, red, green, respectively), where $S=1$.}
\label{fig:werner}
\end{figure}
As shown in Fig.~\ref{fig:werner}, for each $d$,
the solid curve lies above the dashed threshold line when
$-1<x<\frac{2}{d}-1$, indicating that the criterion successfully
detects entanglement in the Werner states. The intersections
$x=\frac{2}{d}-1$ mark the entanglement thresholds, which shift
leftward as $d$ increases, reflecting that higher-dimensional Werner
states are less entangled and thus harder to detect. Each curve has
a cusp at $x=\frac{1}{d}$, where $G=0$. At this point, the Werner
states reduce to the maximally mixed state $\mathbb{I}_{AB}/d^2$,
the unique state with no quantum correlations. The cusp separates
the antisymmetric branch ($x<1/d$) from the symmetric branch
($x>1/d$), marking the transition between the two types of
correlations. At $x=1$, the curve meets the threshold line, which is
consistent with the separability of the Werner states at this
endpoint. Specifically, if we take $d=2$ and $p=\frac{1-2x}{3}$,
the $d$-dimensional Werner states reduce to two-qubit Werner states,
which can be written as
\begin{align}
\notag
\widetilde{\rho}_{AB} = \frac{1-p}{4} \mathbb{I}_{AB} + p \ket{\Psi^-}\bra{\Psi^-}, \quad -\frac{1}{3} \le p \le 1,
\end{align}
where $\ket{\Psi^-}=\frac{1}{\sqrt{2}}(\ket{01}-\bra{10})$. We can
obtain the degenerated criterion: $\widetilde{\rho}_{AB}$ is
entangled if $p>\frac{1}{3}$, which has been given in
\cite{HuangF,RajeevS}.

\vskip0.1in

\noindent {\bf 5. Conclusions}\par We have proposed a measure of
quantum uncertainty to investigate average coherence within the
framework of conical 2-designs GEAM based on metric adjusted skew
information, and demonstrated that it is proportional to the
quantity $Q_f(\rho)$. And we have proven the equivalence of this
measure to the scaled average coherence based on metric adjusted
skew information in the following three frameworks: the unitary
groups, operator orthonormal bases, and MUB. We have also given the
trade-off relations between this measure and quantum $f$ entropy.
Additionally, a trade-off relation exists among the reductions of
this measure to MUB and SIC-POVM, and the maximal coherence of a
state based on metric adjusted skew information, which confirms the
conjecture in \cite{ChenRX}. Furthermore, utilizing this measure and
conical 2-designs GEAM, respectively, we have presented two new
entanglement criteria, and exemplified the effectiveness of them by
isotropic states and $d$-dimensional Werner states, correspondingly.
Since a generalized equiangular measurement can be constructed from
a GEAM via a probability distribution, and since the metric-adjusted
skew information reduces to the GWYD skew information while the
GEAM reduces to scaled MUM, scaled GSIC-POVM, and scaled
$(N,M)$-POVM in specific cases, our results naturally generalize
those found in previous literature. The results in this paper may
shed some new light on the research of average coherence based on
skew information under a set of measurements.

\vskip0.1in

\noindent{\bf Appendix A. Proof of Theorem 1}

Define a monotone metric on the observable space as\cite{PetzD1996}
\begin{align}
\notag K_{\rho}^{f}(A,B) = \frac{f(0)}{2} \mathrm{tr}\left[A^\dagger
\, c_{f}(L_{\rho}, R_{\rho}) \, B\right].
\end{align}
Then we have
\begin{align}
\notag I_{f}(\rho,H) =
K_{\rho}^{f}(\mathrm{i}[\rho,H],\mathrm{i}[\rho,H]).
\end{align}
Note that $K_{\rho}^{f}(A,B)$ is bilinear on $\mathbb{R}$. From the
construction of GEAM, we obtain
\begin{align}
\label{eq19}
C_{f}^{\mathcal{P}}(\rho) =& \frac{1}{N} \sum_{k = 1}^{N} \sum_{l = 1}^{M_k} I_{f}(\rho, P_{k, l}) \notag\\
=&\frac{1}{N} \sum_{k = 1}^{N} \sum_{l = 1}^{M_k} K_{\rho}^{f}( \mathrm{i}[\rho, \frac{a_k}{d} \mathbb{I}_d + \tau_k H_{k, l}], \mathrm{i}[\rho, \frac{a_k}{d} \mathbb{I}_d + \tau_k H_{k, l}] ) \notag\\
=&\frac{1}{N} \sum_{k = 1}^{N} \tau_k^2 \sum_{l = 1}^{M_k} K_{\rho}^{f}( \mathrm{i}[\rho, H_{k, l}], \mathrm{i}[\rho, H_{k, l}] ) \notag\\
=&\frac{1}{N} \sum_{k = 1}^{N} \Bigg[\tau_k^2\sum_{l = 1}^{M_k-1} K_{\rho}^{f}( \mathrm{i}[\rho, H_{k, l}], \mathrm{i}[\rho, H_{k, l}]) \notag\\
&+ \tau_k^2(\sqrt{M}_k+1)^2 K_{\rho}^{f}( \mathrm{i}[\rho, G_k],
\mathrm{i}[\rho, G_k] ) \Bigg].
\end{align}
On the other hand, using the definition $G_k=\sum\limits_{l=1}^{M_k-1}G_{k,l}$, we have
\begin{align}
\label{eq20}
&\sum_{l = 1}^{M_k-1} K_{\rho}^{f}( \mathrm{i}[\rho, H_{k, l}], \mathrm{i}[\rho, H_{k, l}]) \notag\\
=&\sum_{l = 1}^{M_k-1} K_{\rho}^{f}( \mathrm{i}[\rho, G_k-\sqrt{M_k}(\sqrt{M_k}+1)G_{k,l}], \mathrm{i}[\rho, G_k-\sqrt{M_k}(\sqrt{M_k}+1)G_{k,l}]) \notag\\
=& \sum_{l=1}^{M_k-1} \Big[ K_\rho^f(\mathrm{i}[\rho, G_k], \mathrm{i}[\rho, G_k]) - \sqrt{M_k}(\sqrt{M_k}+1) K_\rho^f(\mathrm{i}[\rho, G_k], \mathrm{i}[\rho, G_{k,l}]) \notag \\
&- \sqrt{M_k}(\sqrt{M_k}+1) K_\rho^f(\mathrm{i}[\rho, G_{k,l}], \mathrm{i}[\rho, G_k]) + M_k(\sqrt{M_k}+1)^2 K_\rho^f(\mathrm{i}[\rho, G_{k,l}], \mathrm{i}[\rho, G_{k,l}]) \Big] \notag\\
=&(M_k-1)K_{\rho}^{f}( \mathrm{i}[\rho, G_k], \mathrm{i}[\rho, G_k] ) -2\sqrt{M_k}(\sqrt{M_k}+1)K_{\rho}^{f}( \mathrm{i}[\rho, G_k], \mathrm{i}[\rho, \sum_{l = 1}^{M_k-1}G_{k, l}] ) \notag\\
&+M_k(\sqrt{M_k}+1)^2\sum_{l = 1}^{M_k-1}K_{\rho}^{f}( \mathrm{i}[\rho, G_{k, l}], \mathrm{i}[\rho, G_{k, l}]) \notag\\
=&-(\sqrt{M_k}+1)^2K_{\rho}^{f}( \mathrm{i}[\rho, G_k], \mathrm{i}[\rho, G_k] ) \notag\\
&+M_k(\sqrt{M_k}+1)^2\sum_{l = 1}^{M_k-1}K_{\rho}^{f}(
\mathrm{i}[\rho, G_{k, l}], \mathrm{i}[\rho, G_{k, l}]).
\end{align}
Let $\rho=\sum\limits_{j=1}^{d}\lambda_j\ket{\phi_j}\bra{\phi_j}$ be
the spectral decomposition of $\rho$. Note that
$\{G_0\}\cup\{G_{k,l}\}$ forms an orthonormal Hermitian basis. Then
by combining Eqs. \eqref{eq19} and \eqref{eq20}, and letting $S_k=S$
for all $k=1,\cdots,N$, it holds that
\begin{align}
C_{f}^{\mathcal{P}}(\rho) =& \frac{1}{N}\sum_{k = 1}^{N}\tau_k^2 M_k(\sqrt{M_k}+1)^2\sum_{l = 1}^{M_k-1}K_{\rho}^{f}( \mathrm{i}[\rho, G_{k, l}], \mathrm{i}[\rho, G_{k, l}]) \notag\\
=&\frac{1}{N}\sum_{k = 1}^{N}S_k\sum_{l = 1}^{M_k-1}\frac{f(0)}{2}\sum_{m,n = 1}^{d}(\lambda_m-\lambda_n)^2c_{f}(\lambda_m,\lambda_n)\left|\bra{\phi_m}{G_{k, l}}{\ket{\phi_n}}\right|^2 \notag\\
=&\frac{S}{N}\frac{f(0)}{2}\sum_{m,n = 1}^{d}(\lambda_m-\lambda_n)^2c_{f}(\lambda_m,\lambda_n)\sum_{k = 1}^{N}\sum_{l = 1}^{M_k-1}\left|\bra{\phi_m}{G_{k, l}}{\ket{\phi_n}}\right|^2 \notag\\
=&\frac{S}{N}\frac{f(0)}{2}\sum_{m,n = 1}^{d}(\lambda_m-\lambda_n)^2c_{f}(\lambda_m,\lambda_n)\left(1-\frac{\delta_{mn}}{d}\right) \notag\\
=&\frac{S}{N}Q_{f}(\rho) \notag,
\end{align}
which completes the proof. \qed

\noindent{\bf Appendix B. Proof of Theorem 3}

According to Proposition 5 in \cite{FanY2025}, we have
\begin{align}
\notag I^{\max}_f(\rho)=\frac{1}{d}Q_{f}(\rho).
\end{align}
Let $N=d+1$, $S=\frac{1}{(d+1)^2}$ and $N=1$, $S=\frac{1}{d(d+1)}$
in Eq.\eqref{eq4}, respectively, by Theorem 1 we have
$C_{f}^{\mathrm{MUB}}(\rho)=\frac{Q_{f}(\rho)}{d+1}$ and
$C_{f}^{\mathrm{SIC-POVM}}(\rho)=\frac{Q_{f}(\rho)}{d(d+1)}$, and
thus Eq.\eqref{eq14} follows. \qed

\noindent{\bf Appendix C. Proof of Theorem 4}

Suppose that $\rho_{AB}$ is separable. Then we have
$\rho_{AB}=\sum\limits_{j=1}^{r}p_j \rho_j^A\otimes\rho_j^B$, where
$0\le p_j\le1$, $\sum\limits_{j=1}^{r}p_j=1$, $r$ is the number of
terms in the separable decomposition, and $\rho_j^A(\rho_j^B)$
stands for the density matrix acting on the subsystem $A(B)$, it
holds that
\begin{align}
F_{f}^{\mathcal{P}_A,\mathcal{P}_B}(\rho_{AB})&=\frac{1}{N} \sum_{k = 1}^{N} \sum_{l = 1}^{M_k} I_{f}(\sum\limits_{j=1}^{r}p_j \rho_j^A\otimes\rho_j^B, P_{k, l}^A\otimes \mathbb{I}_B + \mathbb{I}_A\otimes P_{k, l}^B) \notag\\
&\le \frac{1}{N} \sum\limits_{j=1}^{r}p_j \sum_{k = 1}^{N} \sum_{l = 1}^{M_k} I_{f}(\rho_j^A\otimes\rho_j^B, P_{k, l}^A\otimes \mathbb{I}_B + \mathbb{I}_A\otimes P_{k, l}^B) \notag\\
&=\frac{1}{N} \sum\limits_{j=1}^{r}p_j \sum_{k = 1}^{N} \sum_{l = 1}^{M_k} [I_{f}(\rho_j^A,P_{k, l}^A)+I_{f}(\rho_j^B,P_{k, l}^B)] \notag\\
&=\frac{S}{N}\sum\limits_{j=1}^{r}p_j[Q_{f}(\rho_j^A)+Q_{f}(\rho_j^B)] \notag\\
&\le \frac{2S(d-1)}{N}\notag.
\end{align}
The last inequality holds due to the fact that $0\le Q_{f}(\rho)\le
d-1$ for any quantum state $\rho$ \cite{GibiliscoP}. \qed

\noindent{\bf Appendix D. Proof of Theorem 5}

Suppose that $\rho_{AB}$ is separable. Then we have
$\rho_{AB}=\sum\limits_{j=1}^{r}p_j \rho_j^A\otimes\rho_j^B$, where
$0\le p_j\le1$, $\sum\limits_{j=1}^{r}p_j=1$, $r$ is the number of
terms in the separable decomposition, and $\rho_j^A(\rho_j^B)$
stands for the density matrix acting on the subsystem $A(B)$. Hence,
one has $\rho_A=\sum\limits_{j=1}^{r}p_j\rho_j^A$,
$\rho_B=\sum\limits_{j=1}^{r}p_j\rho_j^B$. Thus, in \cite{ZhangCJ},
it holds that
\begin{align}
\notag \rho_{AB}-\rho_A\otimes\rho_B=\frac{1}{2}\left[
\sum\limits_{s,t = 1}^{r}
p_sp_t(\rho_s^A-\rho_t^A)\otimes(\rho_s^B-\rho_t^B) \right].
\end{align}
Using the Cauchy-Schwarz inequality leads to
\begin{align}
&G^{\mathcal{P}^A,\mathcal{P}^B}(\rho_{AB}) \notag\\
\le&\frac{1}{2}\sum_{k = 1}^{N} \sum_{l = 1}^{M_k} \sum\limits_{s,t = 1}^{r} \left| \sqrt{p_sp_t}\mathrm{tr}(P_{k,l}^A(\rho_s^A-\rho_t^A)) \right| \cdot \left| \sqrt{p_sp_t}\mathrm{tr}( P_{k, l}^B(\rho_s^B-\rho_t^B)) \right| \notag\\
\le&\frac{1}{2}\sqrt{\sum_{k = 1}^{N} \sum_{l = 1}^{M_k} \sum\limits_{s,t = 1}^{r}p_sp_t\left[ \mathrm{tr}(P_{k,l}^A(\rho_s^A-\rho_t^A)) \right]^2} \cdot \sqrt{\sum_{k = 1}^{N} \sum_{l = 1}^{M_k} \sum\limits_{s,t = 1}^{r}p_sp_t\left[ \mathrm{tr}(P_{k,l}^B(\rho_s^B-\rho_t^B)) \right]^2} \notag\\
=&\frac{1}{2}\sqrt{2\sum_{k = 1}^{N} \sum_{l = 1}^{M_k}\left\{ \sum\limits_{s = 1}^{r}p_s\left[ \mathrm{tr}(P_{k,l}^A\rho_s^A) \right]^2 - \left[ \mathrm{tr}(P_{k,l}^A\rho_A) \right]^2 \right\}} \notag\\
&\cdot \sqrt{2\sum_{k = 1}^{N} \sum_{l = 1}^{M_k}\left\{ \sum\limits_{t = 1}^{r}p_t\left[ \mathrm{tr}(P_{k,l}^B\rho_t^B) \right]^2 - \left[ \mathrm{tr}(P_{k,l}^B\rho_B) \right]^2 \right\}} \notag\\
\le&S\sqrt{(1-\mathrm{tr}\rho_A^2)(1-\mathrm{tr}\rho_B^2)}\notag,
\end{align}
which finishes the proof. \qed

\noindent

\subsubsection*{CRediT authorship contribution statement}
{\bf Baolong Cheng:} Writing--original draft, Methodology,
Investigation, Formal analysis. {\bf Linlin Ye:} Writing--review \&
editing, Validation. {\bf Zhaoqi Wu:} Writing--review \& editing,
Validation, Conceptualization, Funding acquisition, Supervision.

\subsubsection*{Data availability}
No data was used for the research described in the article.

\subsubsection*{Declaration of competing interest}
The authors declare that they have no known competing financial
interests or personal relationships that could have appeared to
influence the work reported in this paper.

\subsubsection*{Acknowledgements}
This work was supported by National Natural Science Foundation of
China (Grant Nos. 12561084,12161056) and Natural Science Foundation
of Jiangxi Province of China (Grant No. 20232ACB211003).

\end{document}